\documentstyle[amssymb,prl,aps,multicol,epsfig]{revtex}

\begin{document}
\title{Optical conductivity and penetration depth in MgB$_{2}$}
\author{A. V. Pronin$^{1,2}$, A. Pimenov$^{1,}\thanks{%
corresponding author, email:
Andrei.Pimenov@Physik.Uni-Augsburg.DE}$, A. Loidl$^{1}$, and S.
I. Krasnosvobodtsev$^{3}$}
\address{$^{1}$Experimentalphysik V, EKM, Universit\"{a}t Augsburg, 86135 Augsburg,
Germany \\ $^{2}$Institute of General Physics, Russian Academy of Sciences, 119991 Moscow, Russia\\ $^{3}$P. N.
Lebedev Physics Institute, Russian Academy of Sciences, 117924 Moscow, Russia}
\date{\today}
\maketitle

\begin{abstract}
The complex conductivity of a MgB$_{2}$ film has been investigated
in the frequency range 4 cm$^{-1}<\nu <30$ cm$^{-1}$ and for
temperatures 2.7 K $<T<300$ K. The overall temperature dependence
of both components of the complex conductivity is reminiscent of
BCS-type behavior, although a detailed ana\-ly\-sis reveals a
number of discrepancies. No characteristic feature of the
isotropic BCS gap temperature evolution is observed in the
conductivity spectra in the superconducting state. A peak in the
temperature dependence of the real part of the conductivity is
detected for frequencies below 9 cm$^{-1}$. The superconducting
penetration depth follows a $T^{2}$ behavior at low temperatures.
\end{abstract}

%\pacs{74.25.Gz, 74.72.Bk, 74.76.Bz}
\begin{multicols}{2}

Recent discovery of superconductivity at relatively high temperature
 ($T_{\rm c}\approx 39$ K) in a simple binary compound,
magnesium boride \cite{nagamatsu}, has stimulated extensive
theoretical and experimental studies in this material. Most of
these studies are attempting to find the pairing mechanism that
leads to the superconductivity in this compound \cite{hirsch}.
The strongest evidences for phonon-mediated superconductivity in
MgB$_{2}$ comes from the boron-isotope effect measured by the P.
Canfield group \cite{budko}. Further confirmation for such a
mechanism would be an observation of characteristic features
predicted by the BCS-theory for different experimentally measured
quantities, such as an s-wave superconducting gap in tunneling
and optical experiments, or a coherence peak in NMR and in the
real part of the low-frequency conductivity. During several
months passed from the discovery of superconductivity in
MgB$_{2}$, large experimental efforts have been undertaken to
observe such features. However, the absence of high-quality
samples hampers the collection of reliable experimental results.
The tunneling data obtained on ceramic samples \cite{karapetrov}
confirm roughly the BCS predictions for a s-wave gap, although
the ratio $2\Delta (0)/kT_{{\rm c}}$ varies significantly from one
report to another, being between 2.5 to 4. These variations
together with some deviations from the temperature dependence of
the BCS gap have mostly been explained by imperfections of the
sample surface, but alternative explanations e.g. based on
multiple gaps have also been proposed \cite{bascones}. The
photoemission data \cite{takahashi} can be best fitted with an
isotropic s-wave gap using $2\Delta (0)/kT_{{\rm c}}=3$. The NMR
experiments \cite{kotegawa} reveal a tiny coherence peak with
$2\Delta (0)/kT_{{\rm c}}=5$, which indicates that strong
coupling regime dominates in MgB$_{2}$. The temperature dependence
of the penetration depth, which is indicative to the gap symmetry,
shows either quadratic (muon spin rotation experiments
\cite{panagopoulos}) or linear ($H_{{\rm c}2}$ measurements
\cite{li}) laws, both being inconsistent with an activated
behavior, predicted by the BCS-theory. Only one optical study has
been reported till now to our knowledge \cite{gorshunov}. The
superconducting-to-normal grazing reflectivity ratio demonstrates
a gradual increase at frequencies below 70 cm$^{-1}$, which might
be considered as a sign of a superconducting gap with $2\Delta
(0)/kT_{{\rm c}}=2.6$ \cite{gorshunov}. Therefore, the
experimental data are rather controversial, and there is no
general agreement, whether MgB$_{2}$ is a BCS-type superconductor
or not.

In this paper we report on optical investigations of a high
quality MgB$_{2}$ film in the submillimeter frequency range ($4$
cm$^{-1}<\nu <30$ cm$^{-1}$). Both components of the complex
conductivity $\sigma ^{*}(\nu ,T)=\sigma _{1}+i\sigma _{2}$ have
been directly measured as function of temperature and frequency.
This kind of measurements has been already applied to a variety
of high-temperature and conventional superconductors and proven
to be a powerful method for studying the electrodynamical
properties of these materials \cite{pronin,pimenov}.

The MgB$_{2}$ film has been grown by two-beam laser ablation
\cite{nozdrin} on a plane-parallel sapphire substrate $10\times
10$ mm$^{2}$ in size. The crystallographic orientation of the
substrate was $[1\bar{1}02]$ with a thickness of about 0.4 mm.
Magnetization measurements of the film have indicated a sharp
superconducting transition at 32 K with a width of 1 K (upper
frame of Fig. 1). The details of the growth process will be given
elsewhere \cite{prep}. The measurements in the submillimeter-wave
range have been performed using a coherent source spectrometer
\cite{kozlov}. Four backward-wave oscillators (BWO's) have been
employed as monochromatic and continuously tunable sources
covering the range from 4 cm$^{-1}$ to 30 cm$^{-1}$. The
Mach-Zehnder interferometer arrangement has allowed measuring
both the intensity and the phase shift of the wave transmitted
through the MgB$_2$ film on the substrate. Using the Fresnel
optical formulas for the complex transmission coefficient of the
substrate-film system, the complex conductivity has been
determined directly from the measured spectra. The optical
parameters of the bare substrate have been obtained in a separate
experiment. In the present study we have measured the frequency
spectra of the transmittance and of the phase shift at several
temperatures in the normal (above $T_{\rm c}$) and in the
superconducting states of MgB$_2$. In addition, we have also
measured temperature dependences of these quantities at a fixed
frequency (4.1 cm$^{-1}$) continuously from 2.7 to 60 K. A more
detailed description of the experimental setup and of the data
analysis is given in Refs. \cite{pronin,pimenov,kozlov}.

\begin{figure}[tb]
\centering
\includegraphics[width=6cm,clip]{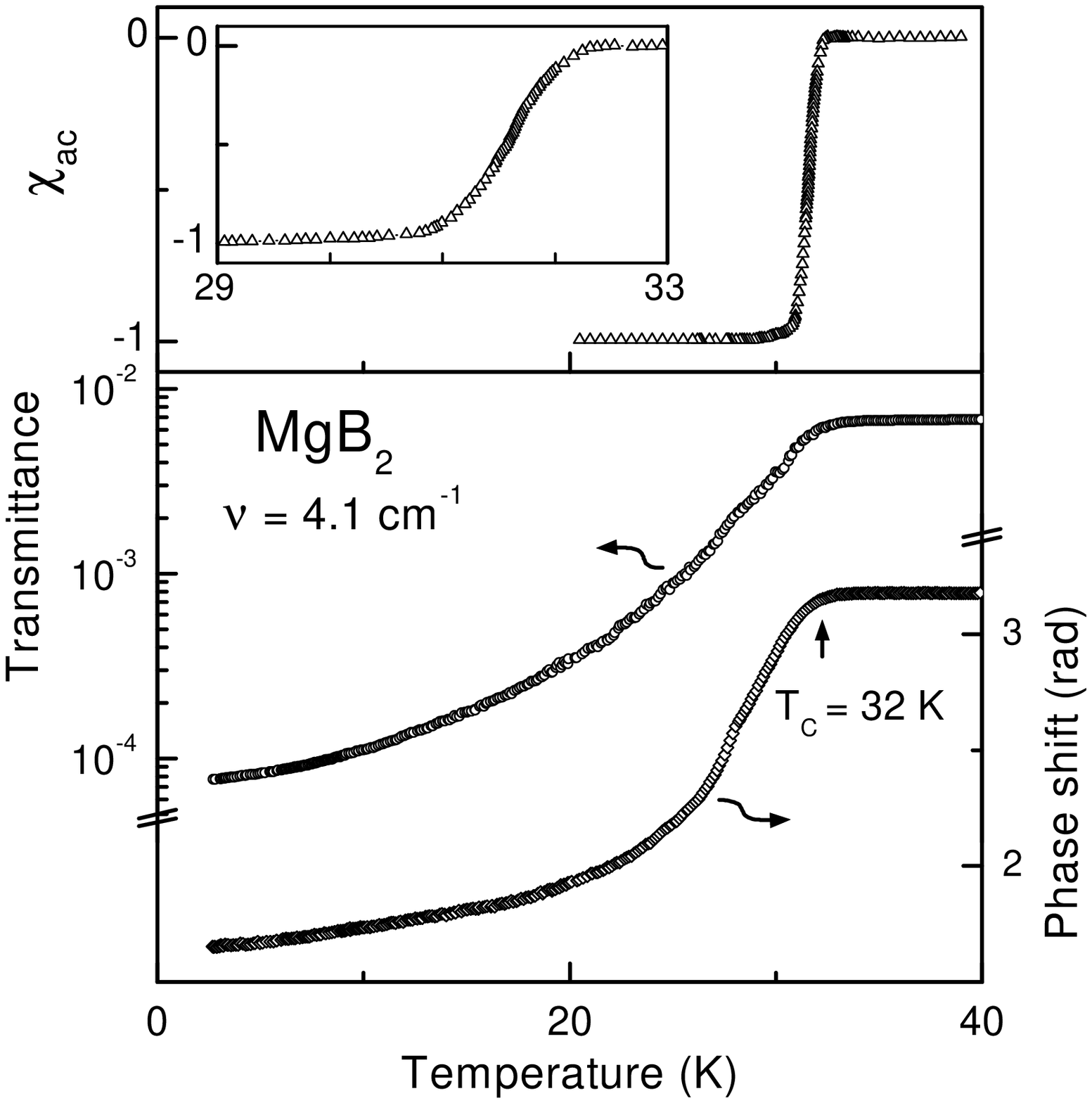}
\vspace{0.2cm} \caption{Upper panel: Temperature dependence of the
magnetic susceptibility of MgB$_{2}$ film on Al$_{2}$O$_{3}$
substrate. Inset shows the data around $T=T_{\rm c}$ on an
enlarged scale. Lower panel: Temperature dependence of the
transmittance at a fixed frequency $\nu =4.1$ cm$^{-1}$ (left
scale) and temperature dependence of the phase shift at the same
frequency (right scale).} \label{fchi}
\end{figure}

The lower frame of Fig. \ref{fchi} represents the temperature
dependence of the transmittance and the phase shift of MgB$_{2}$
film at $\nu =4.1$ cm$^{-1}$. Above the superconducting
transition both quantities are almost temperature independent.
This behavior is consistent with the absence of temperature
dependence of the dc-resistivity in MgB$_{2}$ in this temperature
region \cite{nagamatsu}. The onset of the superconductivity is
immediately reflected in the transmittance and phase shift. Both
start to decrease below $T_{{\rm c}}$, corresponding to an
increase of $\sigma_2$ in the superconducting phase due to
Meissner effect.

\begin{figure}[tb]
\centering
\includegraphics[width=6.5cm,clip]{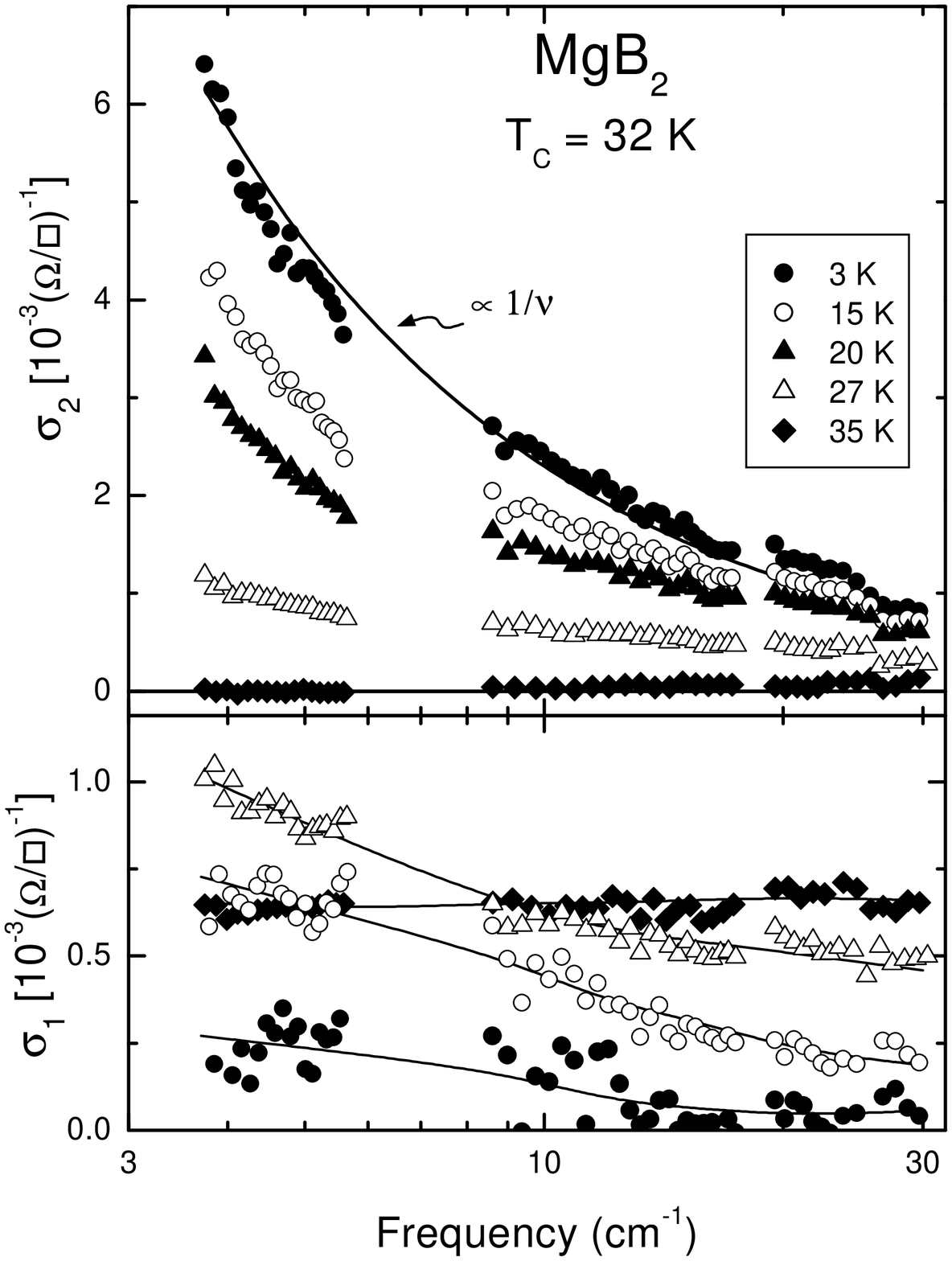}
\vspace{0.2cm} \caption{Frequency dependence of the complex
conductivity of MgB$_{2} $ film above and below $T_{\rm c}$.
Upper panel: imaginary part $\sigma _{2}$. Solid line represents
the $\sigma _{2}\propto 1/\nu $ dependence. Lower panel: real
part $\sigma _{1}$. Lines are drawn to guide the eye. Scattering
of the experimental points and the mismatch between different
BWO's represent the experimental accuracy.} \label{fsigf}
\end{figure}

Examples of the complex conductivity spectra, directly calculated
from the frequency-dependent transmittance and phase shift, are
shown in Fig. \ref{fsigf}. The normal-state conductivity (35 K
curve) demonstrates a typical metallic behavior. The real part of
conductivity is essentially frequency independent, while the
imaginary part is almost zero and exhibits a small linear increase
for increasing frequencies. This indicates a Drude conductivity
with a scattering rate above the measured frequency window.
Simultaneous fitting of the $\sigma_1$ and $\sigma_2$ spectra
with a Drude model gives an estimate for the scattering rate at
35 K: $1/2 \pi \tau = 150^{+70}_{-50}$ cm$^{-1}$.

The transition into the superconducting state gives rise to
significant changes in the spectra. $\sigma_2$ starts to diverge
for $\nu \rightarrow 0$ (Fig. \ref{fsigf}, upper frame). This
divergence increases with decreasing temperature, reflecting a
growth of the spectral weight of the superconducting condensate.
The frequency dependence of $\sigma_2$ can be well described by
the $1/\nu$-dependence, that corresponds to the $\delta$-function
in $\sigma_1$ at $\nu = 0$ via the Kramers-Kronig relations. In
the superconducting state a pronounced frequency dispersion
arises in the $\sigma_1$ spectra as well (Fig. \ref{fsigf}, lower
frame). At low frequencies $(\nu < 9$ cm$^{-1})$ and starting
from the normal state $\sigma_1$ initially increases (by
approximately a factor of 1.5 at the lowest frequency) and then
decreases, while at higher frequencies, $\nu > 9$ cm$^{-1}$,
$\sigma_1$ monotoneously decreases upon cooling. The spectra of
$\sigma_1$ are quite smooth, and no characteristic BCS behavior
with a pronounced minimum at $\nu = 2\Delta(T)$ (a "dip") is
detected. This "dip" is predicted by the BCS theory for a
dirty-limit superconductor with an isotropic s-wave gap at
temperatures between $T = 0$ and $T = T_{\rm c}$ \cite{tinkham}
and has been observed in conductivity spectra of conventional
superconductors (e.g. Ref. \cite{pronin}). The absence of this
feature in the $\sigma_1 (\nu)$ spectra does not allow direct
determination of the superconducting gap value, and might
indicate an unconventional character of the gap parameter in
MgB$_2$. Another possibility would be a fast developing of the
energy gap, which removes the characteristic gap-signature from
the measured spectra already for $T$ close to $T_{\rm c}$.
%or a significant suppression of the scattering rate
%below $T_{\rm c}$. This suppression of the scattering rate would
%put MgB$_2$ into the clean limit, where no peculiarity at $\nu =
%2\Delta(T)$ is expected according to the BCS-theory.
At the lowest
temperatures ($\sim 3$ K) $\sigma_1$ almost approximates zero at
higher frequencies (within the experimental accuracy), indicating
the absence of the normal carrier contribution to the
electromagnetic response at $T \rightarrow 0$, and confirming the
high quality of the film.

\begin{figure}[tb]
\centering
\includegraphics[width=5.5cm,clip]{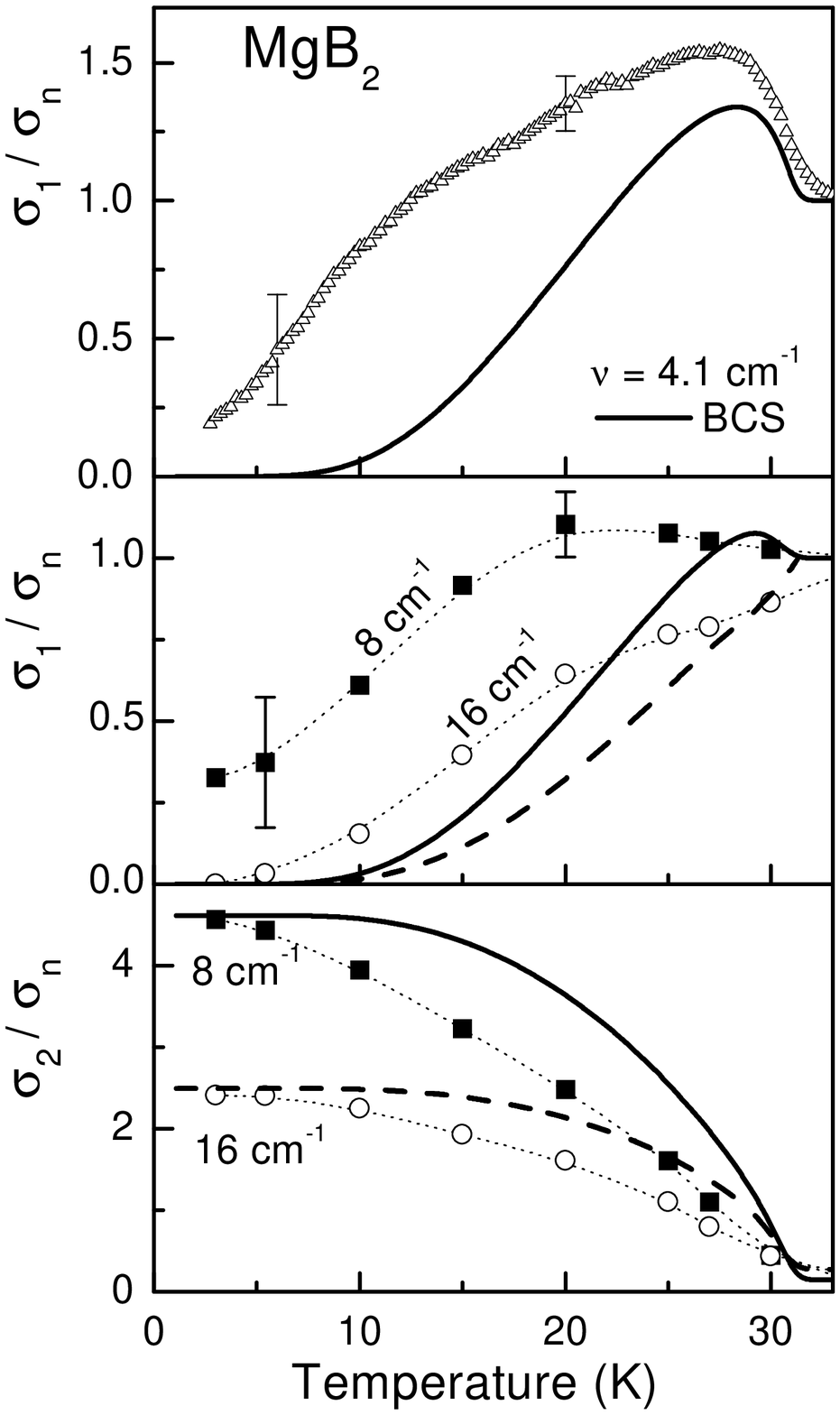}
\vspace{0.2cm} \caption{Temperature dependence of the complex
conductivity of MgB$_{2}$ film for several frequencies. Upper
panel: $\sigma _{1}$ at 4.1 cm$^{-1}$. Solid line represents the
BCS model calculation with $2\Delta /k_{{\rm B}}T_{\rm c}=3.53$,
$1/2\pi\tau = 60$ cm$^{-1}$. Lower panels: temperature
dependencies of the real (middle) and imaginary (bottom) parts of
the complex conductivity at 8.2 cm$^{-1}$ (full squares) and at
16 cm$^{-1}$ (open circles). Thick lines represent the BCS
calculations with the same parameters as in the upper panel,
solid line: $\nu=8.2$ cm$^{-1}$, dashed line $\nu=16$ cm$^{-1}$.
Thin dotted lines are drawn to guide the eye. The error bars are
estimated from the scattering of the experimental points in the
frequency spectra (Fig. \ref{fsigf}).} \label{fsigt}
\end{figure}

From the temperature evolution of the conductivity spectra it
becomes clear that a maximum in the temperature dependence of
$\sigma_1$ should exist below $T_{\rm c}$ and at low frequencies.
The temperature dependencies of $\sigma_1$ and $\sigma_2$ for
several frequencies are shown in Fig. \ref{fsigt}. The temperature
dependence of $\sigma_1$ for $\nu =4.1$ cm$^{-1}$ is calculated
from the transmittance and phase-shift measurements, presented in
Fig \ref{fchi}. (For better representation the data for $\nu
=4.1$ cm$^{-1}$  are shown in a separate frame.) The data in the
two other frames were taken from the frequency spectra (Fig.
\ref{fsigf}). Thick lines in Fig. \ref{fsigt} show the
weak-coupling BCS calculations for the frequencies indicated.
Taking the ratio $2\Delta/kT_{\rm c} =3.53$, the scattering rate
$1/\tau$ remains the only free parameter for these calculations.
We have chosen $1/2 \pi \tau=60$ cm$^{-1}$ that the calculations
at the lowest temperatures match the experimental values of
$\sigma_2$. It has to be noted already at this point that this
scattering rate disagrees with $1/2 \pi \tau = 150$ cm$^{-1}$
obtained from the normal-state conductivity. Taking $1/2 \pi \tau
= 150$ cm$^{-1}$ in the BCS-calculations substantially
overestimates the low-temperature values of $\sigma_2$. The BCS
curves, even with the lowered $1/2 \pi \tau=60$ cm$^{-1}$, do not
perfectly fit to the experimental points, and we were not able to
significantly improve the simultaneous fit of $\sigma_1$ and
$\sigma_2$ data neither by taking different ratios of
$2\Delta/kT_{\rm c}$, nor by introducing a distribution of
superconducting transition temperatures in our calculations.

The peak in the temperature dependence of $\sigma_1$ vanishes at
frequencies above 9 cm$^{-1}$, and the temperature dependence of
$\sigma_1$ at higher frequencies basically follows the behavior
shown in Fig. \ref{fsigt} for $\nu=16$ cm$^{-1}$. The observed
peak in $\sigma_1(T)$ is reminiscent of the coherence peak,
predicted for a BCS superconductor in the dirty limit. However,
the interpretation of this peak in terms of the BCS theory might
contradict to the absence of the characteristic BCS gap signature
in $\sigma_1(\nu)$ spectra discussed above. Both features {\large
(}the coherence peak and the gap in $\sigma_1(\nu)${\large) }
should appear in the dirty BCS limit, and none of them is
expected in the clean limit.

The measured $\sigma_1(T)$ dependencies for all frequencies are
higher then the BCS calculations. This disagreement could be
caused by an additional absorption. This additional absorption
may smear out the BCS gap signature, or may indicate an
unconventional pairing mechanism. It is difficult to say, whether
this absorption is of intrinsic or extrinsic origin. However, the
absorption tends to disappear at $T \rightarrow 0$ (see also the
lower panel of Fig. \ref{fsigf}), which seems to be in favor of
an intrinsic nature of the absorption.

\begin{figure}[tb]
\centering
\includegraphics[width=6cm,clip]{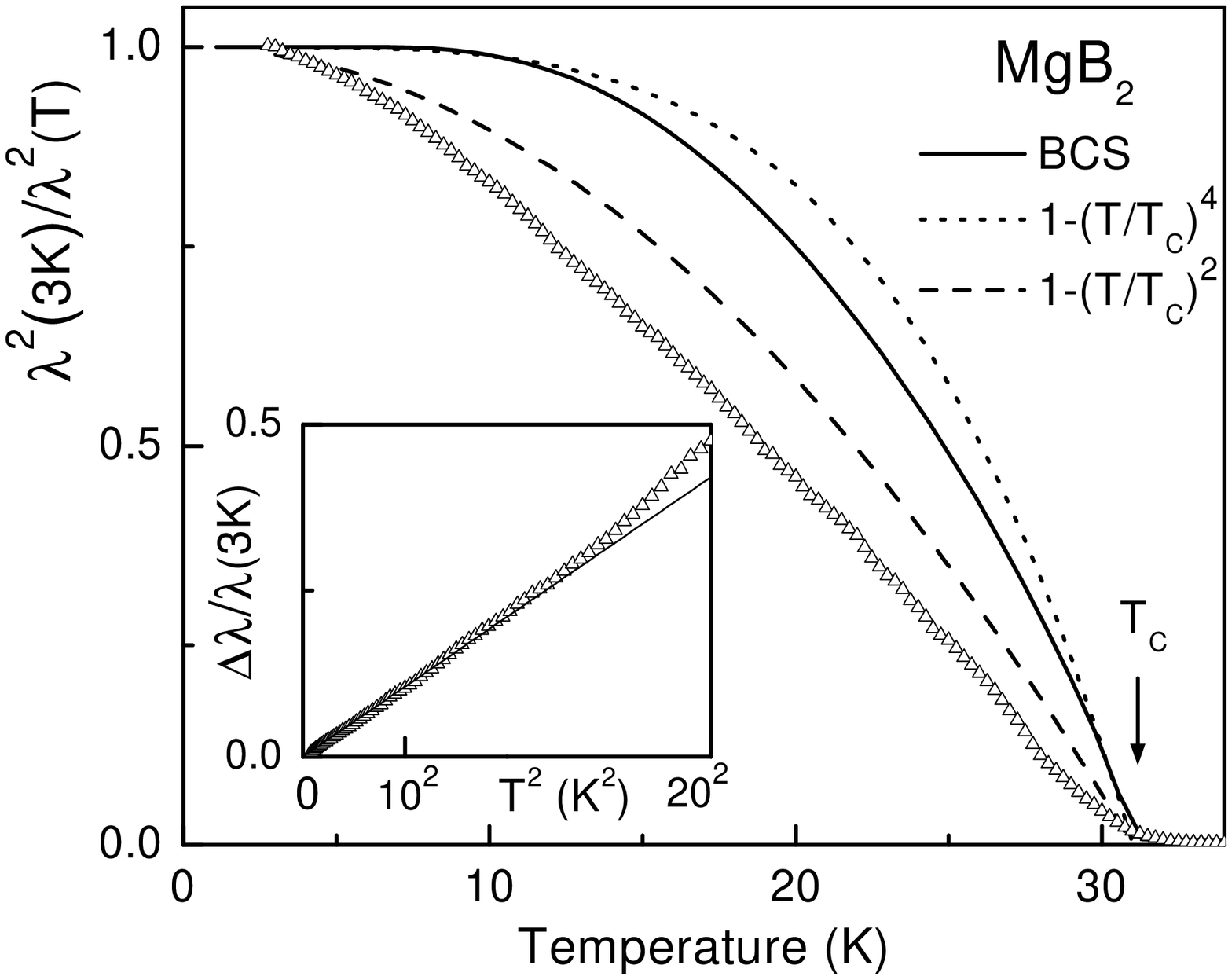}
\vspace{0.2cm} \caption{Temperature dependence of the
low-frequency penetration depth of MgB$_{2}$ film. Symbols:
experimental data as obtained from the temperature-dependent
transmittance and phase shift shown in Fig. \ref{fchi}. Lines
represent various model calculations, solid: weak-coupling BCS,
dotted: two-fluid $[1-(T/T_{{\rm c}})^{4}]$, dashed:
$[1-(T/T_{{\rm c}})^{2}]$. Inset shows the temperature variation
of the penetration depth below 20 K as a function of $T^{2}$.}
\label{flambda}
\end{figure}

The temperature dependence of the imaginary part of the complex
conductivity (Fig. \ref{fsigt}, lower panel) deviates
significantly from the BCS curves. Qualitatively the same
deviations from the BCS calculations have been found for all
measured frequencies: the slope of the experimental $\sigma_2(T)$
curves is more gradual at temperatures just above $T_{\rm c}$,
and at $T \rightarrow 0$ it is steeper then the BCS prediction.
Since the superconducting penetration depth is directly connected
to the imaginary part of conductivity via $ \lambda =(\mu
_{0}\omega \sigma _{2})^{-1/2}$ ($\mu _{0}$ is the vacuum
permeability and $\omega =2\pi \nu $ is the angular frequency),
for the further discussion we focus on the temperature dependence
of the penetration depth (Fig. \ref{flambda}).

Fig. \ref{flambda} compares the penetration depth in MgB$_{2}$
with the predictions of different mo\-de\-ls. The experimental
data (open triangles) have been calculated from the temperature
dependence of the transmittance and phase shift measured at $\nu
=4.1$ cm$^{-1}$ (Fig. \ref{fchi}). The solid line represents the
weak-coupling BCS calculations with the parameters used in Fig.
\ref{fsigt}. The dotted line shows the result of the two-fluid
model $[1-(T/T_{{\rm c}})^{4}]$. The dashed line represents the
$[1-(T/T_{{\rm c}})^{2}]$ dependence. The first two dependencies
reproduce the temperature variation of the penetration depth in
conventional superconductors \cite{tinkham}, and the last one
often reasonably fits the experimental penetration depth in the
high-$T_{{\rm c}}$ cuprates \cite{bonn96}. The experimental data
in Fig. \ref{flambda} clearly deviate from all these model curves
in the whole temperature range below $T_{{\rm c}}$.

The inset of Fig. \ref{flambda} shows the temperature variation
of the penetration depth as a function of $T^{2}$ and at low
temperatures. The experimental points below approximately 15 K
closely follow a straight line, indicating that a $T^{2}$
behavior is a good approximation to the low-temperature data (in
agreement with the muon-spin-rotation measurements
\cite{panagopoulos}). The power-law dependence of the penetration
depth in the zero-temperature limit has been observed in various
high-$T_{{\rm c}}$ superconductors and interpreted as an evidence
for the existence of nodes in the gap function. However, in the
case of a strongly anisotropic s-wave energy gap the transition to
the BCS-like exponential behavior may take place at even lower
temperatures than obtained in the present study ($\sim 2.7$ K).
Therefore, the results of Fig. \ref {flambda} suggest the strong
gap anisotropy, or may even indicate the existence of the nodes
in the gap function.

In conclusion, we have measured the complex conductivity of
MgB$_2$ film in the frequency range 4 cm$^{-1}<\nu \ <30$
cm$^{-1}$ and for temperatures 2.7 K $<T<300$ K. The temperature
dependence of the real part of the low-frequency conductivity
shows a peak below the superconducting transition, which
qualitatively agrees with the predictions of the BCS theory.
However, detailed comparison of the frequency and temperature
behavior of the complex conductivity with the BCS model
calculations shows significant discrepancies between the
experiment and the theory. In addition, we observe no indications
of the temperature evolution of the gap in $\sigma_1 (\nu)$ for
frequencies below 30 cm$^{-1}$. The temperature dependence of the
penetration depth is consistent with the $\delta \lambda \propto
T^2$ behavior below $\sim 15$ K, which implies significant
anisotropy in the gap function.

This work was supported by the BMBF via the contract 13N6917/0 -
EKM. A.V.P. acknowledges partial support by SFB 484.

\end{multicols}
\end{document}